# Contributions from Pilot Projects in Quantum Technology Education as Support Action to Quantum Flagship

S Faletic[1], P Bitzenbauer[2], M Bondani[3], M Chiofalo[4], S R Goorney[5,6], K Krijtenburg-Lewerissa[7], O Mishina[8], R Müller[9], G Pospiech[10], İ Ercan[11], M Malgieri[12], A Merzel[13], M Michelini[14], P Onorato[15], H Pol[16], L Santi[14], Z C Seskir[17], J Sherson[5,6], H K E Stadermann[16], A Stefanel[14], E Surer[18], K Tóth[19], J Yago Malo[20], O Zabello[21]

[1]University of Ljubljana Faculty of Mathematics and Physics, Jadranska ulica 19, 1000 Ljubljana, Slovenia
[2]Friedrich-Alexander-Universität Erlangen-Nürnberg, Department of Physics, Physics Education Research, Staudtstr. 7, 91058 Erlangen, Germany
[3]CNR-Institute for photonics and nanotechnology, via Valleggio 11, Como, Italy
[4]Department of Physics "Enrico Fermi", University of Pisa and INFN-Pisa (Italy)
[5]Department of Management, School of Business and Social Sciences, Aarhus University, Denmark
[6]Niels Bohr Institute, Copenhagen University, Denmark
[7]Freudenthal Institute, Utrecht University, P.O. Box 85170, 3508 AD Utrecht, The Netherlands
[8]Istituto Nazionale di Ottica (CNR-INO), Area Science Park, Basovizza I-34149 Trieste, Italy
[9]Physikdidaktik TU Braunschweig, Bienroder Weg 82, 38106 Braunschweig, Germany
[10]TU Dresden, Department of Physics, Research group physics education, Helmholtzstr. 10, 01069 Dresden, Germany
[11]Delft University of Technology (TU Delft), Department of Microelectronics, Mekelweg 4, 2628 CD Delft, the Netherlands
[12]Department of Physics, University of Pavia, Via Bassi 6 27100 Pavia Italy
[13]The Seymour Fox School of Education, The Hebrew University of Jerusalem, Mount Scopus, Jerusalem 9190501, Israel
[14]PUnità di ricerca in didattica della fisica, Dipartimento di scienze matematiche, informatiche e fisiche, University of Udine, via delle Scienze 206, 33100 Udine, Italy
[15]Physical Science Communication Laboratory, Department of Physics, University of Trento, 38123 Povo, TN, Italy
[16]ELAN, Department of Teacher Development, University of Twente, P.O.Box 217, NL-7500 AE Enschede, The Netherlands
[17]Karlsruhe Institute of Technology, Institute for Technology Assessment and Systems Analysis (ITAS), Karlsruhe, Germany
[18]Department of Modeling and Simulation, Graduate School of Informatics, Middle East Technical University, Ankara, 06800, Turkey
[19]Physics Education PhD Programme, Eötvös Loránd University, Budapest, Hungary
[20]Department of Physics 'Enrico Fermi' & INFN, University of Pisa, Largo B. Pontecorvo 3, I-56127 Pisa, Italy

[21]Offenburg University of Applied Sciences, Badstraße 24, 77652 Offenburg, Germany

sergej.faletic@fmf.uni-lj.si

**Abstract**. The GIREP community on teaching and learning quantum physics and the Education section of the Quantum flagship project of the European Union (QTEdu) have brought together different stakeholders in the field of teaching quantum physics on all levels, including outreach. The goal of QTEdu is to pave the way for the training of the future quantum workforce. To this end, it is necessary to understand the needs of the quantum technology (QT) field, make the general public aware of the existence and importance of QT, and introduce quantum physics already in high school, so that high school students can choose QT as their field of study and career. Finally, new university courses need to be established to support emerging specific profiles such as a "quantum engineer". In this symposium, four QTEdu pilot projects were brought together to demonstrate how their complementary approaches have worked towards realising the above goals.

## 1. Introduction

Quantum technology (QT) is a rapidly developing field which has the potential to become an extensive source of employment with relevant economic impact in the near future. Therefore, there is a growing demand for a quantum workforce. One hindrance for the development of a future quantum workforce is that students often become familiar with quantum physics only at university level, and even that not necessarily in the first year. By then, their career paths may have already been chosen. The goal of the outreach activities described below is, therefore, to bring the quantum world to students before university so they may consider choosing it as their career path.

However, QT education is not only about training skilled workers for occupational fields in quantum technologies. It is also crucial that the population as a whole gains a basic understanding of the underlying physical principles in order to support extensive funding efforts, regulation and adoption of future quantum products. Therefore, it is important to develop and evaluate suitable concepts and approaches for a wide range of recipients, including high school students in formal education. Physics education research has been investigating student conceptions and learning difficulties in quantum physics for decades. In 2019, the different approaches were discussed at a GIREP symposium [1]. A GIREP Thematic Group was formed to discuss ways and means to efficiently communicate quantum physics at pre-university level [1].

In QTs, the industrial applications of basic principles of quantum physics and the general education in quantum physics meet each other: quantum physics leaves the research laboratories and develops novel technological products for society and industry, some of which are still future perspectives, but are nevertheless extensively covered in the media.

To support this development, the European Quantum Technology Flagship was founded [2]. Within the Flagship, Quantum technologies are divided into four areas:

- *Quantum computing.* Single quantum objects (like ions or superconducting states) are used as qubits. Unlike the bits of classical computing, which can be either in state 0 or in state 1, qubits can be brought into quantum mechanical superposition and entangled states. Computations can then be performed in some kind of parallel manner. However, quantum computers are currently expected to only outperform classical computers on very specific tasks.
- *Quantum sensors.* In this field, quantum physical effects are used to measure quantities such as magnetic fields much more precisely than is possible today. This can be used for diagnostic purposes, for example, to measure the magnetic activity of the human brain [3, 4].
- *Quantum communication.* With single photons, one can develop protocols that promise secure ways of communication – a reason to further develop the corresponding technologies.

- *Quantum simulation.* By using one easier-to-control system (i.e., that of cold atoms in optical lattices), one can simulate other quantum mechanical systems, e.g., the behaviour of electrons in ordered systems. This concept, known as *analog quantum simulation* is not universal (unlike quantum computing), but it remains a promising perspective for the development of new chemical catalysts or medical drugs based on molecular simulations, especially in the nearer-term as quantum computers are being developed.

Quantum technologies have the potential to be disruptive not only economically, but also in terms of teaching quantum physics. They enable a new view of quantum physics: the perspective of applications in everyday life and technology. In the language of physics education: Quantum technologies make context-related teaching and learning of quantum physics possible, as is already practised in many other fields of physics.

By being able to focus on concrete applications of quantum physics in practical contexts, quantum physics possibly loses some of its abstractness. In addition, quantum technological applications can be quite motivating: the encryption and decryption of secret messages in quantum cryptography is probably seen by learners as an interesting and highly relevant topic. Moreover, new experimental possibilities are opened up, for example on quantum sensors, on the transmission of information, or on the use of two-state systems as qubits.

In the European Quantum Technology Flagship, the topics of education, outreach and training were pursued within the Education section's QTEdu Coordination and Support Action (CSA) [2] until 2022 and as part of the new Qucats CSA [5] since then. Within QTEdu, 11 different pilot projects on Education, Outreach and Training were established. Four of them were chosen for the 2022 GIREP symposium to discuss the new perspectives on teaching and learning quantum physics arising from the new quantum technologies.The first to react to this new opportunity were outreach programs. The quantum technology context proved a fertile ground to bring quantum physics closer to the general public, because now it had immediate meaning to them. Outreach also has the benefit of not being constrained by formal curricula, which are slow to change.

But, be it formal education programs or outreach, the goal is to achieve quantum literacy: the ability of the general person to understand quantum terminology and the impact of quantum physics on technology and society. To this end, several projects have been launched to develop a modular and as universal as possible assessment instrument that will enable measurement of the effectiveness of each approach.

The contributors to the symposium discussed how outreach and assessment go hand in hand towards the common goal of building the foundation for a quantum workforce.

## 2. Outreach programs

In this section, we reflect on the role of outreach, which we consider to be non-formal education, in the field of QTs. Through inspiration, outreach is a possible route to the long-term development of a quantum-ready workforce, capable of researching, developing, applying, and industrialising quantum computers, sensors, simulators, and communication infrastructure. Increasing efforts on training at the university level [6] are a short-term solution which address the urgent need for a workforce, but do not help to inspire future generations of Quantum Physicists. We believe that outreach serves an important purpose in helping all citizens to navigate and contribute to modern society. Widespread quantum awareness may be of essential importance as the technologies are disseminated into everyday life to overall transform our economy, in many jobs areas such as the legal & ethical [7], financial [8], medical and environmental [9] among many others. These transformations require new patterns of thought and skills among citizens of the public in our present "society of acceleration and uncertainty" [10]. In order to be aware and valuable citizens of the public, we now require many more and different skills than we have in previous times. Exactly which are required, and how we may develop them through outreach, we discuss in the following examples of two projects.

The *Quantum Technologies Education for Everyone* (QuTE4E) project has spearheaded novel research-based approaches to outreach and developed practical guidelines for outreach. The *Italian Quantum Weeks* (IQW) project brought together a wide range of scientists and communicators for a nation-wide coordinated effort that reached thousands of people [11]. Combining knowledge from both projects can bring us a step closer to effective outreach for achieving quantum literacy.

*2.1. Quantum Technologies Education for Everyone*

The pilot project Quantum Technologies Education for Everyone (QuTE4E) [12] has addressed several questions around the what, why, and how of outreach in a research-based approach we call *Physics Outreach Research* (POR). The ultimate aim of QuTE4E has been to develop practical guidelines for conducting outreach activities, given limited time and opportunities available to practitioners.

First we consider the "why" of outreach: for what purpose do we want to communicate Quantum Technologies to the public? We believe that with a suitably designed storytelling, we may both inspire and educate citizens of the public with the skills necessary to navigate the "society of acceleration and uncertainty" in an informal and enjoyable manner that is inclusive, accessible and manageable by the practitioners.

*2.1.1. Outreach for modern thinking skills.* An inspiration for us is the collective works of Howard Gardner, in particular his Five Minds for the Future [13]. Therein, Gardner promotes the value of the disciplined, synthesising, creative, respectful, and ethical minds. We believe that outreach may be an engine to develop these "minds". In particular, Gardner's minds are inherently addressed in the everyday activities of scientists through the so-called inquiry cycle [14]. Thus presenting the public with activities which enable them to make use of inquiry, in a kind of scientific role-play, can enable outreach to have a dual role in promoting quantum awareness and inspiration, as well as helping to equip participants to navigate the "society of acceleration".

Engaging in scientific thought is precisely the disciplinary thinking to which Gardner refers. The aspects of experimentation, conceptualisation, and theory-building develop the synthesising and creative minds [14]. To address the respectful and ethical minds, one must consider *Responsible Research and Innovation* (RRI) [15]. Responsible research and innovation (RRI) in quantum technology. Ethics and Information Technology, 19, 277-294.] which is an essential component of modern science. Significant work within the pilot has gone into developing a narrative which addresses these minds, leading to the development of a storytelling framework Goorney et al. named *culturo-scientific storytelling* (CSS) [16]. The CSS (see figure 1) combines the theoretical framework of the discipline-culture [17] with the inquiry cycle to address Gardner's minds, developing skills such as disciplinary thinking, creativity, and awareness of the ever-changing nature of scientific knowledge.

The CSS was developed for outreach of QT, but it is not specific to it. It could be applied in other rapidly emerging fields, such as Artificial Intelligence. However, what is unique to our field is, perhaps more than any other, the need for creative thought. For example, how can one understand the concept of quantum superposition or entanglement or spin, when no such examples of this exist in daily life? Even more challenging may be those concepts to which the public can relate due to their presence in popular science, such as Quantum Teleportation. QM forces us to "see" what cannot be seen with eyes and experienced with other senses, to connect abstraction and concrete, and to create visualisation forms that would otherwise be unnecessary. The creative mind is of essential importance here, and we believe it to be one of the ways that the CSS can make the greatest impact in our field.

*2.1.2. Community resources - the "how" of outreach.* The CSS provides a theoretical framework for the narrative approach, but it would be abstract without a means through which to implement it. The QuTE4E pilot has been a fertile ground for the development, testing, and validation of resources aiming to support different aspects of the scientific thinking cycle. This has led to the construction of a community toolbox [18], consisting of aids for creativity, experimentation, and formalisation. Use of these tools has proved highly successful, for example in several events during *Italian Quantum Weeks* (see next section), and in *Europe Quantum Day 2022* [19], a large-scale outreach exhibition taking place in the headquarters of the European Commission in Brussels (estimated 15,000 public participants). In these cases, the events would not have been practical to hold without such resources freely available. Other use cases of these tools, such as a series of workshops based around the art-science exhibition *Quantum Jungle* in Pisa [20], and extracurricular courses for high school students [21, 22], have been investigated quantitatively with research instruments, demonstrating the potential value of these tools. Some examples of these include *Quantum Games*, such as *Quantum Moves* [22] and *Quantum Odyssey* [23], both with their own engaging narratives for participants (citizen science and integration with *Qiskit*, respectively). Visualisation tools, such as *Quantum Composer* [24] and the *Quantum Mechanics Visualisation Project* (QuVis) [25], are particularly valuable as a means of fostering creativity whilst remaining grounded in the real workings of quantum mechanics. There are many more such tools available, and they are consistently being developed by groups around the world. Several of these are represented in the pilot, including *ScienceAtHome* [26], *Quarks Interactive* [27], and *QPlayLearn* [28].

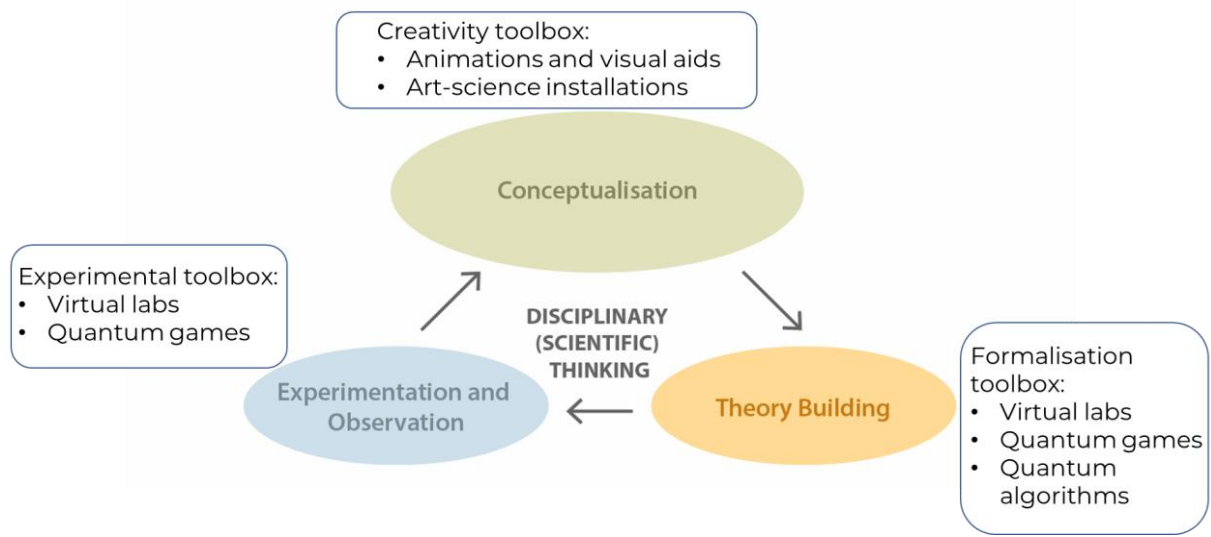

**Figure 1.** The Toolbox of community-developed resources available for quantum technologies outreach, able to support scientific thinking and thus implement the culturo-scientific storytelling. Reproduced from Goorney et al. [16].

Quantum Technologies are complex applications of core concepts in QST. Thus our storytelling, through the CSS, allows us to build these conceptual micro-stories into the macro-story of QT. In so doing, educational paths can be developed on conceptual knowledge in the former case, and on procedural knowledge in the latter. The produced resources could be tailored to the beneficiaries needs, and validated using the work of the QCI pilot described here. Finally, we note that the same approach can be extended from non-formal to formal educational activities, by just changing the level of formal description suited to the audience.

*2.2. Italian Quantum Weeks*
In the framework of the *EU Quantum Flagship*, on the occasion of the first *World Quantum Day* (WQD, April 14, 2022) [29] the Italian scientific and business community have decided to propose the *Italian Quantum Weeks* (IQWs) [30], several weeks from March 26 to May 31 dedicated to the exploration of the foundations and applications of quantum mechanics with the aim to raise awareness of the importance of quantum technologies in our daily lives. This can help promote informed discussions and decisions about how these technologies are developed and used. In addition, teaching quantum technologies to a diverse group of people can help break down barriers and increase engagement with science, which will benefit society as a whole.

More than 130 researchers, technicians, disseminators, communicators, teachers, belonging to more than 40 research institutions, universities, scientific societies, in 17 Italian cities were involved in the project and set up a bundle of initiatives to be proposed to students and general public, with the aim of offering an interactive experience of the peculiar behaviour of the quantum world and of its immense application potential. The activities have been organised with the support of the QTEdu, QTE4E and are officially inserted in the WQD project.

Two national conferences broadcast on the IQWs *YouTube* channel [31]. On April 8, the conference for the general public "From bit to qubit: the advantages of quantum" was held, focusing on the advantages of quantum technologies and the opportunities offered by the Second Quantum Revolution. The conference featured high-level talks by quantum experts from Italy and abroad. Then, the "Quantum@School" conference titled "Stories of Quantum Technologies" took place on April 12. The conference was structured on the talks of five young Italian scientists, who told their personal success stories to inspire students to pursue studies and careers in the quantum field.

Local lectures, guided tours of research laboratories, interactive workshops, quantum game sessions and "Quantum aperitifs" were organised in the different cities, exploiting local expertise and research.

To foster awareness and make connections with personal experience, the "Quantum Suggestions" creativity contest was proposed, aimed at developing a "quantum-inspired" creative project. Participation in the contest is open to people of all nationalities, individually or in groups, and the involvement of schools and students is strongly encouraged. Eligible creative projects may relate to any aspect of quantum mechanics and include any type of artwork. We received 31 artworks of different nature and complexity that were presented to the public during the *European Researchers' Night*. The national winners will be announced at the next *World Quantum Day* on April 14, 2023, and the submitted projects will be included in a national exhibition to be organised in 2024.

The final event of the IQWs 2022 initiatives was held in Palermo on May 22 and consisted of an open conference complemented by a "quantum-inspired" music and dance performance.

*2.2.1. Speaking the Unspeakable.* The most complex and challenging activity of IQWs was the development and organisation of the interactive exhibition "Speaking the Unspeakable-Quantum Superposition," set up in 8 different Italian venues (Catania, Como, Milan, Florence, Modena, Naples, Padua, Rome (two locations)).

The ambitious goal of the exhibition was to introduce the general public to one of the unique properties of quantum states, namely quantum superposition, which, having no classical analogue, is extremely difficult to explain using categories and concepts derived from everyday experience and without resorting to the language of mathematics. The effort of the exhibition was precisely to try to "speak the unspeakable" without misrepresenting the genuine quantum message. To this end, the superposition state was introduced using a quantum experiment that has no classical analogues, namely the single-electron interference experiment proposed by Richard Feynman [32]. A recording of the experiment performed in Italy at the laboratories of the University of Modena and Reggio Emilia was used in the exhibit. In the description of the experiment, we used only the concept of quantum state and decided never to talk about wave-particle dualism so as not to suggest classical analogies that confuse rather than clarify. To support the intuition of the superposition properties of quantum states, we used the analogy with the perception of bistable images, that is, images composed of two different parts that coexist and cannot be seen at the same time.

The heart of the exhibition focused on quantum mechanical description of the world, represented by the axioms of the theory describing preparation, evolution and measurement of quantum states. The probabilistic nature of quantum measurements, the difference between quantum superposition states and classical mixed states, and the noncommutativity of some quantum observables were represented through simple interactive games with coloured cubes, disks and squares of different colours. In addition, the entire logic of the axioms was illustrated with sequences of polarizers.

After discussing the fundamentals of quantum description of the properties of states, the importance of state superposition for applications of quantum technologies, such as computer science and quantum cryptography, was described. First, the concept of "qubit" (minimum amount of quantum information) and "quantum gates" (logical operations on qubits) were introduced, and the advantage of using quantum resources was exemplified through the description and simulation of the classical and quantum "coin-flip" game [33].

Finally, a description of the physical realisation of qubits and the different architectures of quantum computers was presented, along with examples of quantum algorithms implemented on the free IBM *Quantum Experience* platform for quantum computing [34].

The exhibition was supported by written panels, installations, and interactive exhibits. Two original videos on the history of quantum mechanics were placed at the beginning and in the middle of the exhibition, while a video on the experimental research in Italy in the field of quantum technologies closed the exhibition.

*2.2.2. Outcome.* Visitors, both organised school groups and interested individuals, were guided by physics researchers, PhD and University students. The exhibition in the different cities lasted an average of one week and attracted more than 5,000 visitors. Evaluation of the exhibition and other events was carried out in the different locations through a voluntary and anonymous satisfaction questionnaire administered at the end of each event. The questionnaire included items on the level of interest, satisfaction, perceived usefulness, difficulty, and sharing the experience with family members. We obtained 564 responses, 449 from visitors to the exhibition at the different venues and 115 from the other events. The results of a preliminary data analysis are summarised in figure 2 and show that the scores are reasonably high for all items and not too low for the "difficulty" item. figure 2(a) shows that females (51% of participants), males (37%) and those who chose not to answer the question on gender (21%) score similarly, which is quite good considering the known difficulties in involving girls in STEM activities. In contrast, figure 2(b) shows that teachers and common visitors, who are likely to be interested and educated, score better than students in all categories, a result that indicates the need to increase student literacy and awareness by teaching quantum physics in high schools. Finally, figure 2(c) shows that the different types of in-person activities scored the same, while online conferences were much less effective, probably due to a different level of personal involvement and a gradual disengagement with online activities due to the saturation of the use of this mode during the COVID-19 pandemic.

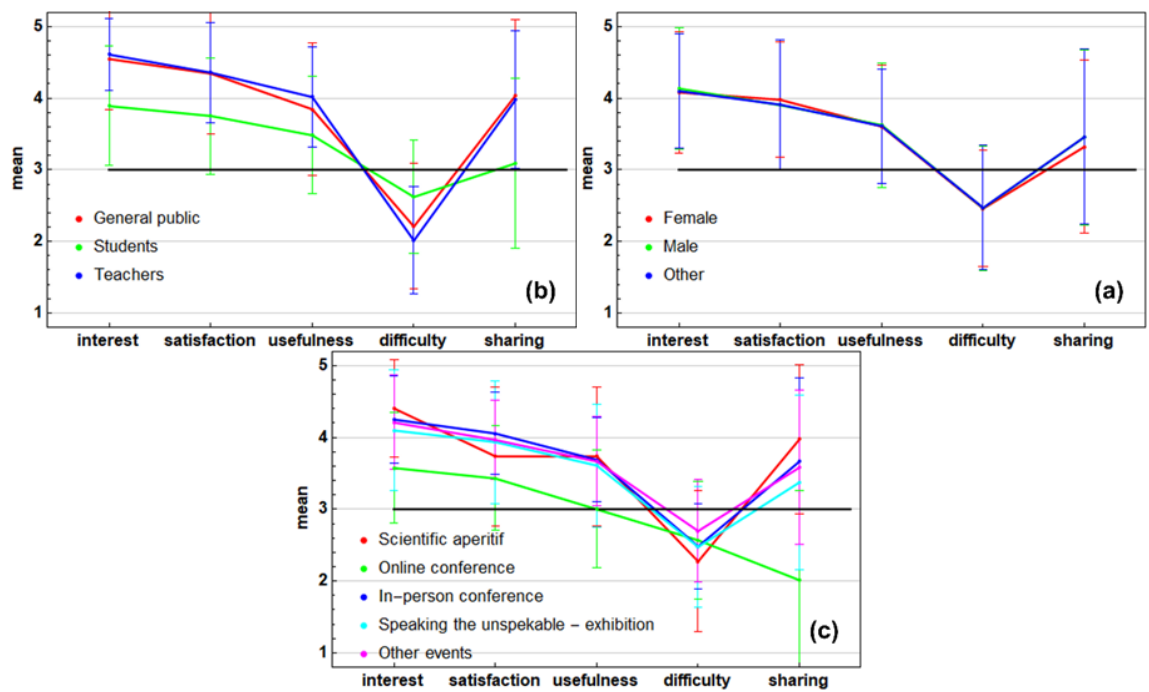

**Figure 2.** Results of the satisfaction questionnaire administered at the end of IQWs outreach events. The panels show the mean score value according to the different items: interest in the activity, satisfaction, perceived usefulness, difficulty, and level of sharing the experience with family and friends. Panel (a) compares the results by gender, panel (b) by category of participants, and finally panel (c) by type of event.

## 3. Assessment

The two examples of outreach presented in this paper do not stand alone. Over the last decades, teaching proposals for quantum physics taking multiple approaches (e.g., history-based, experiment-based, two-state approaches amongst others) with a focus on different key topics [35] have been developed and implemented to make quantum physics (QP) accessible both for the secondary school level (for example, see [36–44]) and the general public. To be able to evaluate and refine the developed teaching and outreach proposals, it is important to have instruments available which can be used to assess students' conceptual understanding within a broad spectrum of approaches, but applicable to the framework of the individual approach. Since there is no such instrument available [45], two initiatives for developing such an instrument were started within the QTEdu pilot projects.

The first initiative is the project *Community-based development of the Quantum Concept Inventory (QCI)*. The aim of this project is to develop a modular quantum concept inventory, the QCI, based on community input which allows for the assessment of students' understanding of quantum physics' key concepts in different contexts. The QCI will be useful for evaluating the numerous teaching concepts and for comparing their impact on student learning of quantum physics despite their different emphasis. Hence, the QCI will serve as an important resource for the quantum physics education research community with regards to robust empirical research into teaching and learning of quantum physics.

The second initiative is the project *Development of quantum concepts via different two-state approaches* (DQC-2stap). The ultimate goal of the project DQC-2stap is to give recommendations for secondary school teachers or educators at university on how two-state approaches and appropriate teaching material can be optimised in the sense that students best can get a grasp of the quantum concepts. To this end, steps have been taken to develop a questionnaire that is intended to be an assessment module that can be administered to any student of any two-state approach in any setting with only minor modifications to concretize particular questions to the context of the approach.

*3.1. The Quantum concept inventory*

The QCI project [46] aims to create a concept inventory that is based on the communities' perspectives and that is applicable for different contexts and key topics. The four main research objectives in this project are:

*3.1.1. Project objectives.* To create a concept inventory that is based on the communities' perspectives and that is applicable for different contexts and key topics, we pursue four main research objectives in this project:
1. Identification of quantum physics' key concepts to be taught to secondary school students from a community perspective,
2. Collection of existing test instruments to create a platform of existing test items on the key concepts of quantum physics,
3. Creation of additional test items in various task formats (e.g., multiple choice, likert scale items, open-ended questions, concept cartoons), and
4. Evaluation of test items, both qualitatively (e.g., think-aloud interviews) and quantitatively in a later stage of the project.

*3.1.2. Current status of the project.* So far, the project members have collected test instruments published in the literature to assess students' understanding of quantum physics topics and have categorised the existing items in pillars with respect to the content aspects covered in an online community event (research objective 2). This categorization helps to identify content domains for which there is:
- a sufficient set of empirically tested items that allow for the assessment of learners' conceptual understanding at secondary school level;
- a lack of empirically evaluated test items to assess students' understanding.

From our categorization, it became apparent that:

- most of the items published in the literature cover the quantum formalism ([47, 48]), topics associated with wave-particle dualism (cf. [49, 50]), matter waves (cf. [51]) or time evolution (cf. [52]).
- Most of the items published in the literature have been developed for the use with university students and, hence, are predominantly not suitable for the use in the secondary school setting (cf. [53]).
- There are no items available in the literature yet that are suitable for evaluating secondary school students' understanding of quantum concepts which are seen as particularly relevant with respect to understanding modern applications of quantum technologies (for example superposition, entanglement, or measurement).

To close the above-mentioned gap of empirically validated test items suitable for the secondary school setting, it is important to understand which content domains are considered relevant for teaching and learning in this educational context today – and more importantly – in the future. Hence, in the sense of research objective 1, the authors decided to conduct a Delphi study to identify the key concepts that are foreseen to be relevant in future quantum physics curricula in secondary schools across Europe, and that consequently should be covered in the QCI. This procedure ensures content validity of our instrument in an early stage of its development.

*3.1.3. The future quantum physics curriculum at secondary schools: A Delphi study.* A Delphi study is currently being conducted to identify key concepts for secondary school teaching of QP that represent that community-based view. In the first round of this Delphi study, physicists, physics teachers, and physics education researchers were asked why secondary school students should learn quantum physics and what topics should be addressed in classroom teaching. In the second round, a questionnaire was distributed, in which the QP community was asked to select which of the topics that emerged from round 1 (including various specifications considering QTEdu's Competence Framework [54]) should be part of the secondary school curriculum (see figure 3).

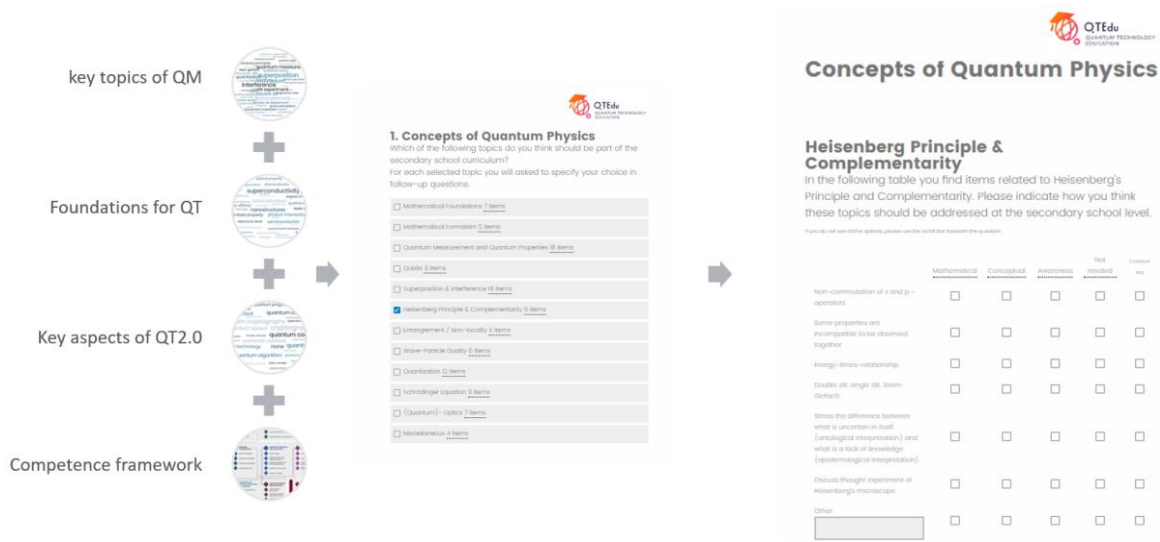

**Figure 3.** The experts' answers in round 1 questionnaire were categorised in three pillars "key topics of QM", "foundations for QT" and "key aspects of QT 2.0" and were, together with input from the European Competence Framework for Quantum Technologies, used to develop the questionnaire of Delphi round 2. In this round, the experts were asked to rate whether the topics provided should be part of future secondary school curricula on which level (not needed, awareness, conceptual, mathematical).

Lastly, a third round is planned: The most frequently selected topics will be ranked by a group of selected experts. Additionally, focus group discussions will help to derive consensus among the experts and will, finally, lead to an adopted version of the European Competence Framework for Quantum Technologies with respect to the secondary school level.

*3.1.4. Outlook.* Based on the results of the Delphi study, the QP topics that should be addressed by the QCI will be determined. Subsequently, new test items will be developed (research objective 3) before the QCI will be evaluated empirically, both using qualitative and quantitative methods (research objective 4). An overview of the QCI development is provided in figure 4.

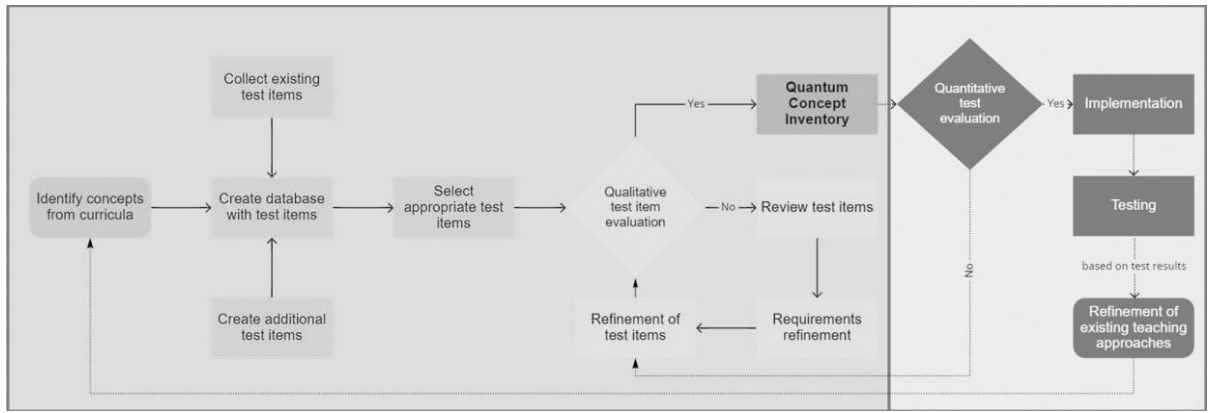

**Figure 4.** Overview of the development of the QCI.

*3.2. Development of quantum concepts via different two-state approaches*
A two-state approach to quantum physics seems to be a suitable and flexible means for teaching and learning quantum physics. In two-state systems systems not only are the basic quantum concepts clearly visible but also discoverable by students themselves while a comprehensible mathematical formalism can be used. Such an approach can be taught in different contexts, e.g. with polarisation, spin, a double-well system or with still other systems with two basis states, including several instances of quantum technology.

However, it could be that different approaches have different strengths and weaknesses when it comes to teaching the most important quantum concepts such as quantum state, superposition, uncertainty, entanglement and measuring process. In order to find hints in this direction in the project *Development of quantum concepts via different two-state approaches* (DQC-2stap) [55] a questionnaire is being developed that sheds light onto this question. This process is accompanied by corresponding teacher education activities.

The questionnaire is designed to be useful within different approaches and settings. It will also help identify which of the two-state contexts is most suitable for teaching which quantum physics concept at the secondary school level. Especially specific learning difficulties of students should be identified.

The gained insights into students' learning difficulties in two-state approaches will be used to develop or improve secondary school or outreach teaching proposals and material based on the two-state approach within a design based research process. This research goal needs more time and will profit from cooperation with other projects in the QTEdu initiative.

*3.2.1. Development of questionnaire and preliminary results.* In order to limit the complexity of the questionnaire development, we initially restricted ourselves to the concept of measurement in quantum physics. This concept was also chosen because the measurement process lies at the core of the peculiarity of quantum physics. So the research question is: "Which understanding and learning difficulties do students show with respect to the measuring process?"

The first version of the questionnaire was developed with the help of existing materials and our own suggestions which have been graded individually by each member of the team, discussed and then approved or rejected by a consensus. The items were mostly composed of two tiers: closed single choice items and an explanation of the choice. These were supplemented by a few open items.

In the next step, the questionnaire was tried with a group of students, who agreed to participate in the study, and adapted and refined based on their answers. Then a pilot with teacher students was done. The resulting final version was administered to students of existing two-state-approach courses in Hungary, Slovenia and Italy. The results were analysed with a coding scheme developed iteratively from the data. The codes were selected to represent students' ideas about the measurement process in quantum mechanics. Preliminary results seem to indicate statistically significant differences between the courses both in the correctness of responses and the distribution of codes associated with the responses. Further analysis is under way towards achieving the goal of identifying differences between courses and students' ideas about measurement as they emerge in different courses.

High school teachers were actively involved in the testing phase of the questionnaire. A teacher training course was developed and conducted in Udine (Italy). The training course consisted of 5 meetings (4 hours each, once a week) and was completed in March 2022 preparing teachers for the implementation of a field study into teaching/learning quantum physics via a polarisation approach (one of the two-state approaches included in our pilot project). The teachers attending the course implemented the two-state course in their classes, collected data on the questionnaire and contributed in the coding discussion. More teacher training courses are intended to be developed in the project, benefitting from the results of the research conducted with questionnaires.

*3.2.2. Outlook.* Using the questionnaire, we will conduct field studies in order to compare learning outcomes from students who are introduced to measurement in quantum physics via different two-state approaches. We will use the outcomes of the questionnaire for a) a refinement of the different teaching sequences, and b) for answering the question about the suitability of different two-state-approaches to introduce students to measurement in quantum mechanics. Should the instrument prove valid, reliable and easily implementable, it could be part or an inspiration for the measurement module of the QCI.

## 4. Discussion and conclusion

The pilot projects launched under the *European Quantum Flagship Educational Coordination and Support Action* aim to address all aspects of making the general public aware of quantum physics and technologies and prepare the ground for a future quantum workforce. This consists of identifying core concepts to teach, developing methods to successfully communicate the concepts to the recipient, and evaluating their understanding of said concepts.

The QCI project with its Delphi study provided basic concepts that experts in the field deem fundamental. These concepts need to be efficiently communicated to the general public in a timely manner to achieve quantum literacy for responsible decision-making about quantum technologies.

Outreach, as non-formal education, can reach large numbers of people and develop in its recipients the minds of aware and capable citizens of the future who understand that scientific ideas can evolve and popular beliefs can change. For this reason, it is essential that non-formal education is engaging and rigorous, independently of the recipients' age and the educational context.

The QUTE4E pilot has designed an original, research-based approach, (*Physics Outreach Research*) and theoretically framed their guidelines in the *culturo-scientific story-telling* [8], which resulted in a toolbox of innovative educational experiences [18, 20, 22–28], which aim to complement the limited experimental and mathematical literacy typical of outreach participants.

The IQW project was a massive initiative by more than 130 content creators and thousands of visitors aiming to educate the general public about the quantum world and quantum technologies. Many types of outreach activities were prepared. The *Speaking the unspeakable* exhibition endeavoured to present the concept of superposition to the general public. In the spirit of physics outreach research, it collected surveys from participants, the results of which show that the topics are equally scored by both genders, that people with more background knowledge consider the topics less difficult and that in-person activities were scored better than on-line ones.

To assess how well participants learn in the various formal and informal programs, a standardised instrument to measure quantum concept knowledge is needed. The QCI project approaches this with ambitious goals. A Delphi study is currently underway to determine the topics to be evaluated. The next steps will develop a modular instrument applicable to all pre-university quantum physics programs. The DQC2-stap project starts with only two-state approaches and only the topic of measurement, but adds teacher education based on its findings. If successful, it could provide a proven stepping stone for the QCI.

The four presentations of the symposium thus cohere very well in the overarching goal of the QTEdu of educating the public about quantum technologies enough for them to consider a career in quantum engineering. Moreover, while significant steps forward have been realised with the QTEdu efforts to qualify and foster non-formal education on quantum technology, we believe that a long way is still to go and the presented pilots may be catalysts to enhance physics education research and accelerate physics outreach research.

**Acknowledgements**
We would like to thank all members of all pilots, listed in web pages of the respective projects [12, 30, 46, 55] for the dedicated work that they put into each pilot. MB would especially like to thank Italo Testa for his help in the creation and analysis of the questionnaire.